\documentclass[sigconf,nonacm]{acmart}

\AtBeginDocument{%
  \providecommand\BibTeX{{%
    \normalfont B\kern-0.5em{\scshape i\kern-0.25em b}\kern-0.8em\TeX}}}

\usepackage{multirow}
\usepackage{tabularray}
\usepackage{braket}
\usepackage{soul}

\begin{document}

\title[]{Impact of Error Rate Misreporting on Resource Allocation in Multi-tenant Quantum Computing and Defense}


\author{Subrata Das}
\email{sjd6366@psu.edu}
\affiliation{%
  \institution{The Pennsylvania State University}
  \city{State College}
  \state{Pennsylvania}
  \country{USA}
  \postcode{16801}
}

\author{Swaroop Ghosh}
\email{szg212@psu.edu}
\affiliation{%
  \institution{The Pennsylvania State University}
  \city{State College}
  \state{Pennsylvania}
  \country{USA}
}


\begin{abstract}
Cloud-based quantum service providers allow multiple users to run programs on shared hardware concurrently to maximize resource utilization and minimize operational costs. This multi-tenant computing (MTC) model 
relies on the error parameters of the hardware for fair qubit allocation and scheduling, as error-prone qubits can degrade computational accuracy asymmetrically for users sharing the hardware. To maintain low error rates, quantum providers perform periodic hardware calibration, often relying on third-party calibration services. If an adversary within this calibration service misreports error rates, the allocator can be misled into making suboptimal decisions even when the physical hardware remains unchanged. We demonstrate such an attack model in which an adversary strategically misreports qubit error rates to reduce hardware throughput, and probability of successful trial (PST) for two previously proposed allocation frameworks, i.e. Greedy and Community-Based Dynamic Allocation Partitioning (COMDAP) \cite{niu2023enabling, upadhyay2024share}. Experimental results show that adversarial misreporting increases execution latency by 24\% and reduces PST by 7.8\%. We also propose to identify inconsistencies in reported error rates by analyzing statistical deviations in error rates across calibration cycles. 

\end{abstract}

%






\maketitle

\section{Introduction}
\label{sec:intro}

Quantum computing is emerging as a powerful tool for solving complex problems in optimization, cryptography, and material science \cite{nielsen2010quantum, castelvecchi2017ibm}. It is becoming more accessible through cloud-based platforms. These platforms allow multiple users to submit programs that execute on shared quantum hardware \cite{castelvecchi2017ibm, niu2023enabling}. Efficient hardware allocation is important for ensuring high throughput and fair resource distribution. However, these systems rely on calibrated error rates of the hardware to guide allocation strategies \cite{niu2023enabling,upadhyay2024share}. If an adversary manipulates these reported values during calibration, resource allocation can become inefficient in terms of execution time and hardware utilization. 

Error rates in quantum computers are dynamic. They change due to environmental conditions, calibration drift, and crosstalk effects \cite{nielsen2010quantum,tannu2019not,kjaergaard2020superconducting}. 
To manage these variations, cloud providers use periodic calibration \cite{magesan2012efficient,kjaergaard2020superconducting,preskill2018quantum, arute2019quantum} either in-house \cite{arute2019quantum} or via third-party services. The latter choice arises mainly due to the high cost and operational complexity associated with frequent calibration. 
These specialized external services have expertise in optimizing calibration procedures, which can improve hardware performance. Outsourcing also allows quantum providers to focus on infrastructure development rather than frequent recalibration. Additionally, third-party calibration provides an independent assessment of hardware performance. Since these services operate separately from the quantum provider, they can offer unbiased verification of reported error rates. This reduces the risk of systematic biases in calibration and ensures that error metrics accurately reflect the hardware’s true performance. 
However, if a malicious third party falsifies the calibration data and misreports error rates, this may misguide the allocation framework to make suboptimal decisions. The adversary does not need to alter the actual hardware noise. Simply by misreporting error parameters, they can mislead the allocator to degrade quantum computation performance.

\begin{figure}[h]
  \centering
  \includegraphics[width=\linewidth]{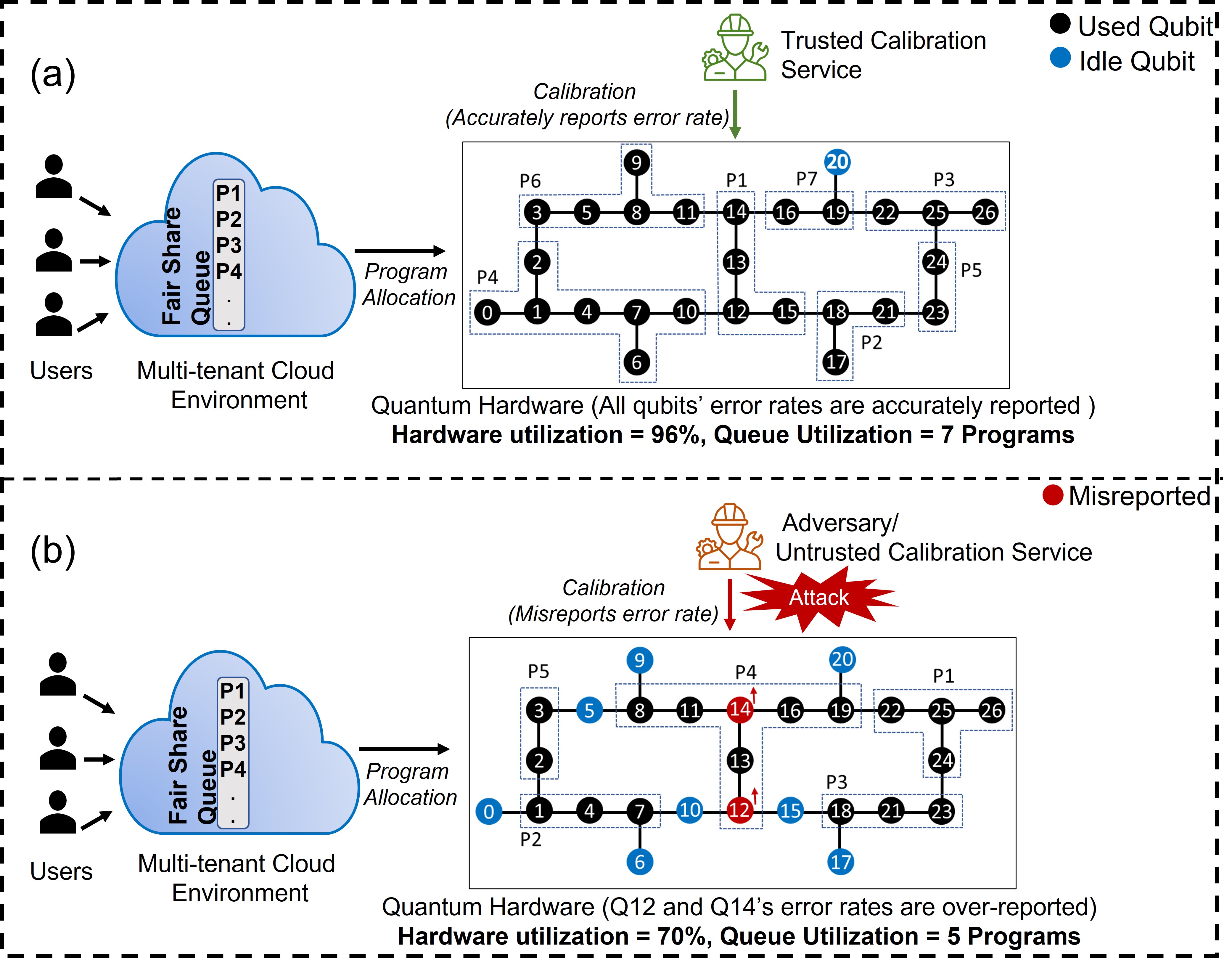}
  \caption{Overview of the proposed threat model. (a) In the absence of an attack, a trusted calibration service accurately reports error rates of all qubits, the allocator assigns qubits efficiently and achieves high hardware/queue utilization. (b) Under attack, the adversary in an untrusted calibration service selectively over-reports error rates for two high-connectivity qubits, Q12 and Q14. This results in inefficient partitioning and lower hardware/queue utilization.}
  \label{fig:attackmodel}
\end{figure}

Specifically, a malicious adversary can use this strategy to increase the number of scheduling rounds required to complete a given number of queued programs. In a multi-tenant computing (MTC) setup, multiple users submit jobs simultaneously. The scheduler groups these jobs into batches based on qubit availability. If the adversary degrades efficient allocation, more rounds will be needed to execute the same workload. This will subsequently delay job completion time, leading to higher execution latency. Fig. \ref{fig:attackmodel} illustrates the idea further. Another attack goal is to increase the depth and gate count of the compiled circuits. When a circuit is mapped to physical qubits, the compiler selects an initial layout that minimizes costly operations, such as SWAP gates. If error rates are misreported, the compiler may prioritize or avoid certain qubits unnecessarily. This leads to additional SWAP gates and longer gate sequences. As a result, the probability of successful trial (PST) will decrease and the noise accumulation will increase. The impact is especially severe in noisy intermediate-scale quantum (NISQ) devices, where error rates are already high.  

Existing quantum computing allocation frameworks optimize for throughput or fidelity. Greedy allocation \cite{niu2023enabling, liu2024qucloud+} method makes local optimization decisions based on reported error rates. More advanced frameworks, such as Community-Based Dynamic Allocation Partitioning (COMDAP) hl{\cite{upadhyay2024share}, use community detection algorithms to form hardware partitions. 

\textbf{Contributions:}
We (a) propose an attack model where an adversary misreports error rates to degrade hardware throughput in multi-tenant quantum computing, (b) evaluate and compare the impact of this attack on two different allocation frameworks, i.e. Greedy and COMDAP, (c) suggest a detection framework to identify error misreporting.

\textbf{Paper Organization:} Section \ref{sec:background} provides background on MTC based quantum computing and existing allocation frameworks. We also review the related work. Section \ref{sec:threat} defines the threat model and adversarial capabilities. Section \ref{sec:heuristics} presents two adversarial heuristics and evaluates their impact on Greedy and COMDAP allocation frameworks. Section \ref{sec:defense} proposes a defense and Section \ref{sec:conclusions} concludes the paper with key findings.

\section{Background}
\label{sec:background}

\subsection{Quantum Circuits and Compilation} 

Quantum circuits use quantum gates to manipulate qubits \cite{nielsen2010quantum}. A qubit exists in a superposition of basis states $\ket{0}$ and $\ket{1}$, represented as: $\ket{\psi} = \alpha_0 \ket{0} + \alpha_1 \ket{1}$, where $\alpha_0$ and $\alpha_1$ are complex numbers satisfying $|\alpha_0|^2 + |\alpha_1|^2 = 1$. Multi-qubit states form a superposition over $2^n$ basis states. Quantum gates apply unitary transformations to qubits. Single-qubit gates include the Pauli-X gate, which flips $\ket{0} \leftrightarrow \ket{1}$, and the Hadamard gate, which creates superposition. The Controlled-NOT (CNOT) gate entangles two qubits by flipping the target qubit based on the control qubit. Quantum circuit output is determined by measurement operation. Entanglement and superposition
are two main properties of quantum processors that enable the
execution of powerful quantum algorithms.

Quantum circuits must be compiled before execution that involves gate decomposition, qubit mapping, and SWAP insertion \cite{nielsen2010quantum}. High-level operations are decomposed into hardware-supported gates and logical qubits are mapped to physical qubits while respecting hardware connectivity. If a two-qubit gate involves non-adjacent qubits, SWAP gates reposition them.
Superconducting quantum hardware suffers from connectivity constraints \cite{preskill2018quantum}. A coupling graph represents valid qubit interactions where each node shows
a qubit and the edges show entanglement capabilities. A CNOT gate can be applied only if a direct edge exists between qubits. Missing connections require SWAP gates, increasing circuit depth. Each SWAP decomposes into three CNOT gates. Therefore, more SWAPs lead to higher noise accumulation and lower fidelity.

\subsection{Multi-tenant Quantum Computing (MTC)} 

Quantum cloud platforms provide shared access to quantum hardware and allows multiple users to run programs remotely. Companies like IBM, Google, and AWS Braket offer these services, eliminating the need for users to maintain costly quantum infrastructure \cite{niu2023enabling, castelvecchi2017ibm}. However, efficient resource allocation and scheduling remain major challenges. Most current cloud-based quantum systems execute one program at a time, leaving many qubits idle. This results in significant resource underutilization. MTC addresses this by scheduling multiple programs to run simultaneously. 

High demand for quantum computing also creates execution bottlenecks \cite{niu2023enabling, liu2024qucloud+}. Long queues delay computation, especially for iterative algorithms requiring multiple executions. Effective scheduling in MTC can reduce such latency and improve system throughput. Schedulers allocate qubits based on system constraints, using strategies like fidelity-aware allocation and fair-share policies \cite{ravi2021quantum}. They optimize qubit selection while considering connectivity, error rates, and circuit depth to ensure reliable execution. Fair scheduling prevents monopolization by prioritizing jobs from users with lower past usage. Platforms like IBM Quantum implement these strategies to distribute computational opportunities equitably across multiple users.

\subsection{Quantum Hardware Allocation Frameworks}

MTC requires efficient resource allocation to maximize hardware utilization and fidelity. The allocation framework decides which qubits are assigned to each program, considering constraints such as connectivity and error rates. Two widely used frameworks are Attractor Node-Based Greedy Allocation \cite{niu2023enabling, liu2024qucloud+} and Community-Based Dynamic Allocation Partitioning (COMDAP) \cite{upadhyay2024share}. The greedy method prioritizes immediate optimization based on individual qubit properties, while COMDAP considers global hardware structure to improve long-term fairness and efficiency.

\textbf{Greedy:} The Greedy approach selects qubits based on the Composite Fidelity Metric, $\text{CFM} = d + (1 - (\text{E} + \text{R}))$
where \( d \) is the degree of the qubit, \(\text{E}\) is the average CNOT error, and \(\text{R}\) is the readout error \cite{niu2023enabling}.
The allocation starts by identifying an attractor node with the highest CFM score. A breadth-first search then expands the partition by adding neighboring qubits with the best scores. The process stops when the partition reaches the required size. This method ensures high qubit utilization but does not balance resource distribution across programs. It can also lead to fragmentation when multiple high-scoring qubits are allocated without considering overall hardware structure.

\textbf{COMDAP:} COMDAP takes a different approach by structuring the quantum hardware into logical partitions before allocation. It applies the Louvain Community Detection Algorithm to form clusters based on connectivity and error rates. These clusters define potential allocations that maintain coherence among selected qubits. The framework then evaluates partitions using the Connectivity and Reliability Index, $\text{CRI} = \frac{\frac{D_{\text{partition}}}{C_{\text{partition}}} + \alpha (1 - (E_{\text{partition}} + R_{\text{partition}}))}{\frac{D_{\text{hardware}}}{C_{\text{hardware}}} + \alpha (1 - (E_{\text{hardware}} + R_{\text{hardware}}))}$, where \( D \) (Density) is the ratio of actual to possible edges, \( C \) (Compactness) is the ratio of observed diameter to maximum possible diameter, and \( \alpha \) is a weighting factor, set to 1 \cite{upadhyay2024share}.
If a cluster exactly matches the program size, it is assigned. Otherwise, the densest subset is chosen, or smaller communities are merged to meet the requirements. This method improves allocation fairness and minimizes unnecessary SWAP operations.

\subsection{Quantum Hardware Calibration}
Quantum error rates fluctuate over time due to temperature variations, electromagnetic interference, crosstalk, and quantum-state relaxation \cite{kjaergaard2020superconducting}. To mitigate these effects, quantum hardware providers perform periodic calibration \cite{arute2019quantum, chen2018metrology}.
The frequency of calibration depends on hardware stability. For example, IBM calibrates superconducting qubits every 12 to 24 hours \cite{acharya2020lightweight}. Calibration process typically includes the following steps:


\textbf{Qubit Control Parameters Calibration:} Qubits require precise calibration to ensure high-fidelity state transitions. Each qubit has a distinct resonance frequency, which drifts over time due to environmental noise and temperature variations. Frequency calibration is performed by sweeping microwave drive frequencies and measuring qubit response to identify and lock onto the optimal resonance \cite{arute2019quantum}. Amplitude calibration is achieved through \textit{Rabi oscillation} experiments, where drive pulses of varying amplitudes are applied, and the qubit's transition probability is measured. The system selects the amplitude that achieves full population transfer between $\ket{0}$ and $\ket{1}$ \cite{chen2018metrology}. Additionally, pulse shape optimization is employed to suppress errors arising from higher-order energy levels and unwanted qubit interactions. Techniques like \textit{drag pulse shaping} mitigate leakage and reduce phase errors \cite{werninghaus2021leakage}.

\textbf{Gate Error Characterization and Benchmarking:} After calibrating control pulses, quantum hardware providers benchmark the performance of quantum gates to assess their fidelity. The error rate is typically measured using \textit{Randomized Benchmarking} (RB), where sequences of randomly chosen Clifford gates are applied, followed by their inverse, and the fidelity decay is analyzed to estimate gate error \cite{magesan2012efficient}. In this process, single-qubit gates are first benchmarked to estimate coherence-limited errors. Once single-qubit performance is optimized, two-qubit gates such as CNOT and CZ are benchmarked, as they introduce additional sources of error due to qubit-qubit interactions and coupling mechanisms. These errors include crosstalk, residual entanglement, and parasitic interactions, which degrade gate fidelity. To mitigate these effects, quantum hardware providers employ cross-resonance gate calibration, echo pulses, composite sequences, and active cancellation methods to suppress unwanted interactions and optimize two-qubit gate performance \cite{patterson2019calibration}.


\textbf{Readout and Decoherence Error Estimation:} Measurement errors arise due to imperfections in qubit-state discrimination, leading to incorrect readout classification. To estimate readout errors, qubits are prepared in known $\ket{0}$ and $\ket{1}$ states and repeatedly measured to determine the probability of misclassification, which is used to construct a readout error matrix \cite{arute2019quantum, chen2018metrology}. Additionally, decoherence effects must be continuously monitored and mitigated. T\textsubscript{1} relaxation time, which quantifies how long a qubit remains in its excited state before decaying to the ground state, is measured by preparing the qubit in $\ket{1}$ and tracking its decay profile. T\textsubscript{2} dephasing time, which defines how long a qubit maintains phase coherence in a superposition state, is typically measured using the Ramsey experiment, where a qubit is placed in a superposition and allowed to evolve freely before applying a second pulse to measure phase coherence decay, providing T\textsubscript{2}*. To account for low-frequency noise, the Hahn spin-echo experiment is used, where an additional refocusing pulse compensates for slow environmental fluctuations, yielding the corrected T\textsubscript{2} value \cite{arute2019quantum}. By continuously updating system parameters and optimizing control pulses, the calibration process ensures that quantum hardware remains operationally stable despite environmental fluctuations. 




\subsection{Related Work}
Quantum hardware allocation plays an important role in MTC for ensuring efficient execution of quantum circuits while minimizing errors. Various approaches have been explored to optimize qubit allocation and mitigate noise-induced performance degradation. However, these methods largely assume trusted error reports, leaving systems vulnerable to adversarial manipulation.

Early works in quantum circuit mapping focused on optimizing execution by minimizing gate errors and ensuring efficient qubit connectivity. Siraichi et al. \cite{siraichi2018qubit} formalized the qubit allocation problem and demonstrated that optimal mappings are NP-complete, necessitating heuristic solutions. Subsequent research introduced error-aware mapping techniques, incorporating physical qubit constraints to enhance fidelity \cite{tannu2019not}. Murali et al. \cite{murali2019noise} proposed noise-adaptive compiler mappings that dynamically adjust qubit assignments based on error rates. These works established the foundational principles of error-aware compilation but did not consider the risks of adversarial error misreporting. 

More recent studies have highlighted the role of error parameters in quantum circuit execution. Acharya and Saeed \cite{acharya2020lightweight} analyzed how manipulated error rates affect single-circuit compilation and mapping but did not consider multi-tenant execution. In contrast, the SHARE framework \cite{upadhyay2024share} explored fair resource allocation in multi-tenant quantum computing, incorporating error parameters and crosstalk effects. However, it did not investigate active threat models based on intentionally misreported calibration data. Our work extends these studies by examining how adversarial misreporting impacts multi-tenant quantum computing, specifically in qubit allocation strategies. 




\section{Threat Model and Adversary Capability}
\label{sec:threat}
\subsection{Threat Model}

The proposed threat model assumes that the quantum provider outsources hardware calibration to a third-party service to reduce operational costs and leverage specialized expertise. The adversary operates within this third-party service. They follow all calibration steps correctly i.e., tuning qubit frequencies, optimizing control pulses, and executing standard error characterization techniques like RB. However, they report slightly modified error data to the quantum service provider (without changing the actual hardware noise). 
By over-reporting errors on well-connected, low-noise qubits and under-reporting those on unreliable qubits, they misguide the allocator into making suboptimal assignments. 
CNOT error rates typically fluctuate significantly over time \cite{tannu2019not}. We assume that the adversary manipulates error rates within the bounds of natural temporal variations, making detection difficult. 

The primary objective of the adversary is to sabotage the efficiency of quantum resource allocation during multi-tenant computing. This can be achieved through two specific attack goals:

\textbf{Reducing hardware throughput:} The attacker misreports error rates to increase the number of execution rounds required to complete a certain number of programs in the queue. A typical multi-tenant quantum queue may contain hundreds to thousands of programs. If poor allocation results in higher idle qubits per round, overall hardware utilization will drop. This will lead to longer wait times and increased execution costs for all users.

\textbf{Reducing the probability of successful trials}: The adversary also aims to degrade fidelity. This is achieved by forcing inefficient compilation requiring excessive SWAP operations. This increases circuit depth, accumulates noise, and lowers the likelihood of obtaining correct computational results.

\subsection{Adversary Capability}

We make the following assumptions regarding the capabilities of the adversary.

\textbf{Ability to misreport Gate Error rate:} We assume that the adversary has access to the calibration data but cannot directly introduce physical noise into the hardware. It is also assumed that the adversary can only modify reported values for gate errors but cannot alter calibration data for readout error, $T_1$, or $T_2$. This assumption is made because CNOT error rates are easier to misreport as compared to the other types of errors. CNOT error rates are measured using randomized benchmarking, which relies on statistical fitting \cite{magesan2012efficient}. By misreporting the number of applied Clifford gates or slightly adjusting the decay curve, the adversary can subtly alter error rates. In contrast, readout error, $T_1$, and $T_2$ require direct physical measurements, making them harder to falsify without detection.

\textbf{Knowledge of coupling maps and allocation frameworks:} The adversary has access to the hardware’s coupling graph and has sufficient computational resources to scrutinize them. He/she also
understands the allocators' decision-making process, whether it follows a greedy approach or an advanced method like COMDAP. These insights can be gained from prior experiments with the allocation framework. They can analyze how different frameworks prioritize qubits based on connectivity, error rates, and density. Most allocation strategies rely on common factors such as qubit connectivity, error rates, and spatial density. For example, Greedy allocators prioritize immediate low-error qubits \cite{liu2024qucloud+, niu2023enabling}, while COMDAP groups qubits into spatially coherent clusters to reduce SWAP operations \cite{upadhyay2024share}. An adversary can leverage this knowledge to strategically misreport values in a way that systematically degrades performance across different frameworks. 





\textbf{Stealth constraints:} To remain undetected, the adversary cannot introduce drastic anomalies in reported error rates. Instead, they make subtle modifications within expected temporal fluctuations to avoid triggering immediate anomaly detection mechanisms. This will be further discussed in section \ref{sec:defense}.


\section{Adversarial Heuristics for Misreporting and Experimental Validation}
\label{sec:heuristics}
\subsection{Overview}
This section presents two adversarial heuristics designed to maximize the impact of misreporting error rates in quantum hardware allocation. The first heuristic aims to increase the number of execution rounds required to finish a given number of jobs in the queue. 
The second heuristic targets to reduce PST by increasing transpiled circuit depth. To evaluate the effectiveness of these heuristics, we conduct experiments using a variety of quantum circuits such as $grover\_n2$, $linearsolver\_n3$, $simon\_n6$, $adder\_n10$ etc. sourced from \cite{li2023qasmbench}. These benchmarks encompass a range of qubit counts (2-12), gate counts (4-1000) and circuit depths. We simulate adversarial misreporting on the Fake27QPulseV1 backend (27 qubit noisy simulator mimicking IBM's Hanoi system) using Qiskit \cite{javadi2024quantum}.

\begin{figure}[h]
  \centering
  \includegraphics[width=\linewidth]{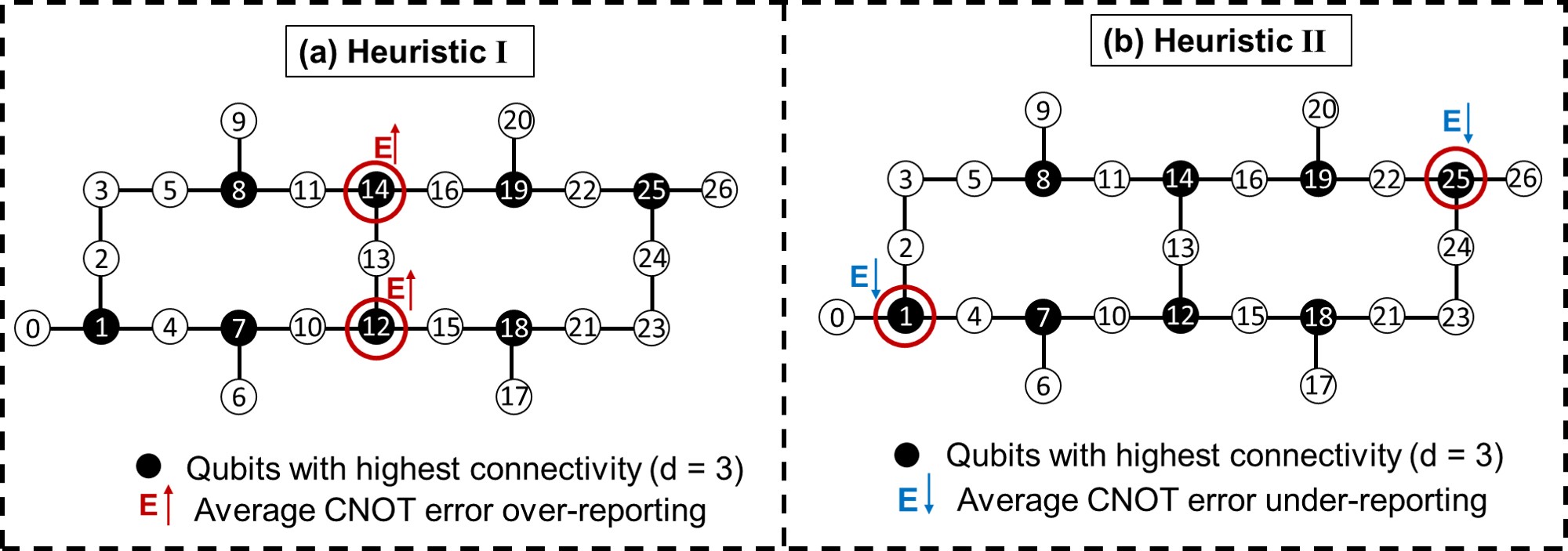}
  \caption{Proposed heuristics. (a) Heuristic I: Over-reporting central high connectivity qubits; (b) Heuristic II: Under-reporting distributed high-Connectivity qubits }
  \label{fig:heuristic}
\end{figure}

\subsection{Heuristic I: Over-Reporting Central High Connectivity Qubits}

\subsubsection{Objective} The adversary aims to reduce hardware throughput by manipulating the allocator to avoid key qubits.

\subsubsection{Methodology} The adversary selectively over-reports error rates for qubits with both high-connectivity and central location. These qubits serve as key bridges in the coupling graph. Over-reporting their error rates might mislead the allocator, forcing fragmented subgraphs, and lowering hardware utilization. To systematically identify target qubits, the adversary follows these steps:

1. \textit{Identify High-Connectivity Qubits:} A qubit $q$ is considered high-connectivity if its degree $d(q)$ in the coupling graph is among the highest in the system. The adversary computes $d(q)$ for all qubits:
   \begin{equation}
   d(q) = \sum_{q' \in N(q)} 1,
   \end{equation}
   where $N(q)$ is the set of qubits directly connected to $q$. All qubits with the highest $d(q)$ are enlisted.

2. \textit{Determine Central Qubits:} The adversary evaluates the spread of shortest-path distances for each high-connectivity qubit. A qubit with low variance in its distance distribution is centrally located. The standard deviation $\sigma_q$ of shortest-path distances is:
   \begin{equation}
   \sigma_q = \sqrt{\frac{1}{|Q|} \sum_{q' \in Q} (p(q, q') - \mu_q)^2},
   \end{equation}
   where $p(q, q')$ is the shortest-path distance between $q$ and $q'$, and $\mu_q$ is the mean distance. The adversary selects top $n$ qubits with the lowest $\sigma_q$. 
   
3. \textit{Over-Report Error Rates:} The adversary inflates the reported error rates of the selected qubits by $k\%$ to force the allocator to avoid them.  $n$ and $k$ values depend on the hardware size and qubit connectivity. The adversary selects the smallest $n$ that maximally disrupts allocation while choosing a $k$ that causes inefficiency without immediate detection (i.e., stays within the range of temporal variation).

Fig. \ref{fig:heuristic} (a) illustrates this strategy on Fake27QPulseV1 backend. Qubits 1, 7, 8, 12, 14, 18, 19, 25 have the highest connectivity (d = 3). Among them, qubits 12 and 14 are centrally positioned as their distance distributions have the lowest variance. Following this heuristic, the adversary will over-report their error rates to achieve the first attack goal.

\subsubsection{Experimental Validation}

\begin{figure}[h]
  \centering
  \includegraphics[width=\linewidth]{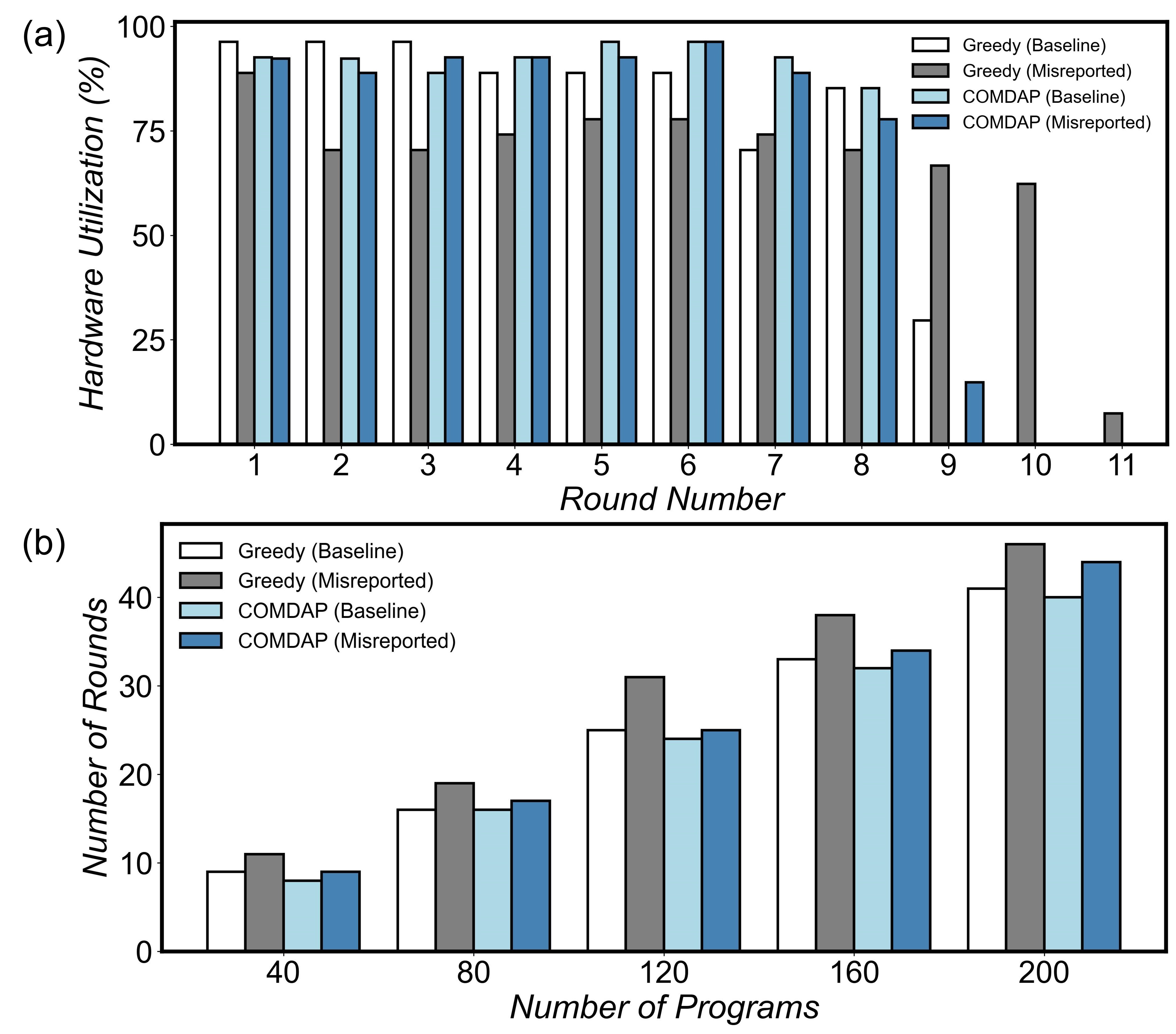}
  \caption{Impact of attack using Heuristic I on hardware throughput. (a) Hardware utilization percentage across rounds for executing 40 queued programs, (b) Total number of rounds required to complete different number of queued programs under baseline and misreported conditions.}
  \label{fig:fig3}
\end{figure}

To evaluate the impact of attack using Heuristic I, we simulated adversarial misreporting on Qiskit's Fake27QPulseV1 backend. The coupling graph contains eight high-connectivity qubits with degree $d=3$ (Fig. \ref{fig:heuristic}). From this set, we selected three centrally located qubits ($Q7, Q8, Q12$) using Eqs. (1) and (2). We experimented with higher values of 
$n$ and $k$, attacking more qubits and inflating errors beyond 15\%. However, the results did not show a significant deviation in allocation inefficiency. Therefore, following the stealth constraints outlined in Section 3.2, we set $n=3$ and $k=15\%$ as reasonable parameters to maximize allocation disruption while avoiding detection.

We executed 40 quantum programs (each requiring 2–10 qubits) under two conditions: (1) baseline allocation using true error rates and (2) adversarial allocation with misreported rates. To assess the impact of the attack, we measured hardware utilization, defined as the percentage of active qubits per round. Fig. \ref{fig:fig3}(a) illustrates how misreporting affects utilization across execution rounds. Under baseline conditions, both Greedy and COMDAP allocators achieve high utilization by efficiently distributing qubits. However, with adversarial misreporting, Greedy allocation suffers a significant decline, with utilization dropping by up to 26\%. The allocator struggles to form valid partitions as central high-connectivity qubits are avoided.  COMDAP, being a community-based approach, maintains slightly better performance but still experiences reduced utilization (as much as 10\%) due to forced fragmentation. 

Fig. \ref{fig:fig3}(b) examines the impact on execution rounds required to process different numbers of queued programs (40 to 200). Under attack, the number of rounds increases significantly, especially for Greedy allocation, which is more sensitive to individual qubit quality. Over-reporting central high-connectivity qubits leads to 24\% longer execution times and reduced throughput. 

\subsection{Heuristic II: Under-Reporting Distributed High-Connectivity Qubits}

\subsubsection{Objective} 
The adversary aims to reduce PST by increasing transpiled circuit depth and noise accumulation. 

\subsubsection{Methodology}
The adversary selectively under-reports the error rates of high-connectivity qubits in a distributed manner. In MTC, typically the allocators prioritize high-connectivity qubits with low error rates during program allocation. Hence, if the adversary attacks following this heuristic, it may manipulate the allocator into selecting distant high-connectivity qubits for a single program. This will force suboptimal partitioning and increase compiled circuit depth, SWAP gate count, and ultimately reduce PST.The adversary follows these steps:

1. \textit{Select a Primary Target Qubit:} The heuristic starts with identifying all high-connectivity qubits in the coupling graph. Among the identified high-connectivity qubits, the adversary selects one qubit $q_1$ and under-reports its error rate by $k_1\%$.

2. \textit{Expand to Distant High-Connectivity Qubits:} The adversary selects the high-connectivity qubit $q_2$ that is farthest from $q_1$ in terms of shortest-path distance $p(q_1, q_2)$ and under-reports its error rate by $k_2\%$. 

3. \textit{Further Distribute Targets:} The process is repeated iteratively. The next target qubit $q_i$ is selected such that it maximizes the minimum path-distance from all previously selected qubits:
   \begin{equation}
   q_i = \arg\max_{q \in H} \min_{q' \in Q_t} p(q, q'),
   \end{equation}
   where $H$ is the set of high-connectivity qubits and $Q_t$ is the set of already selected qubits. Each new target is under-reported by $k_i\%$, where $k_1 > k_2 > k_3 > ... > k_n$.

 Fig. \ref{fig:heuristic} (b) illustrates this heuristic on the Fake27QPulseV1 backend. The adversary first selects qubit 1 and under-reports its error. Next, it selects qubit 25, which is the farthest high-connectivity qubit from qubit 1, and misreports its error at a lower rate. This pattern continues to achieve the second attack goal.


\subsubsection{Experimental Validation}
\begin{figure}[h]
  \centering
  \includegraphics[width=\linewidth]{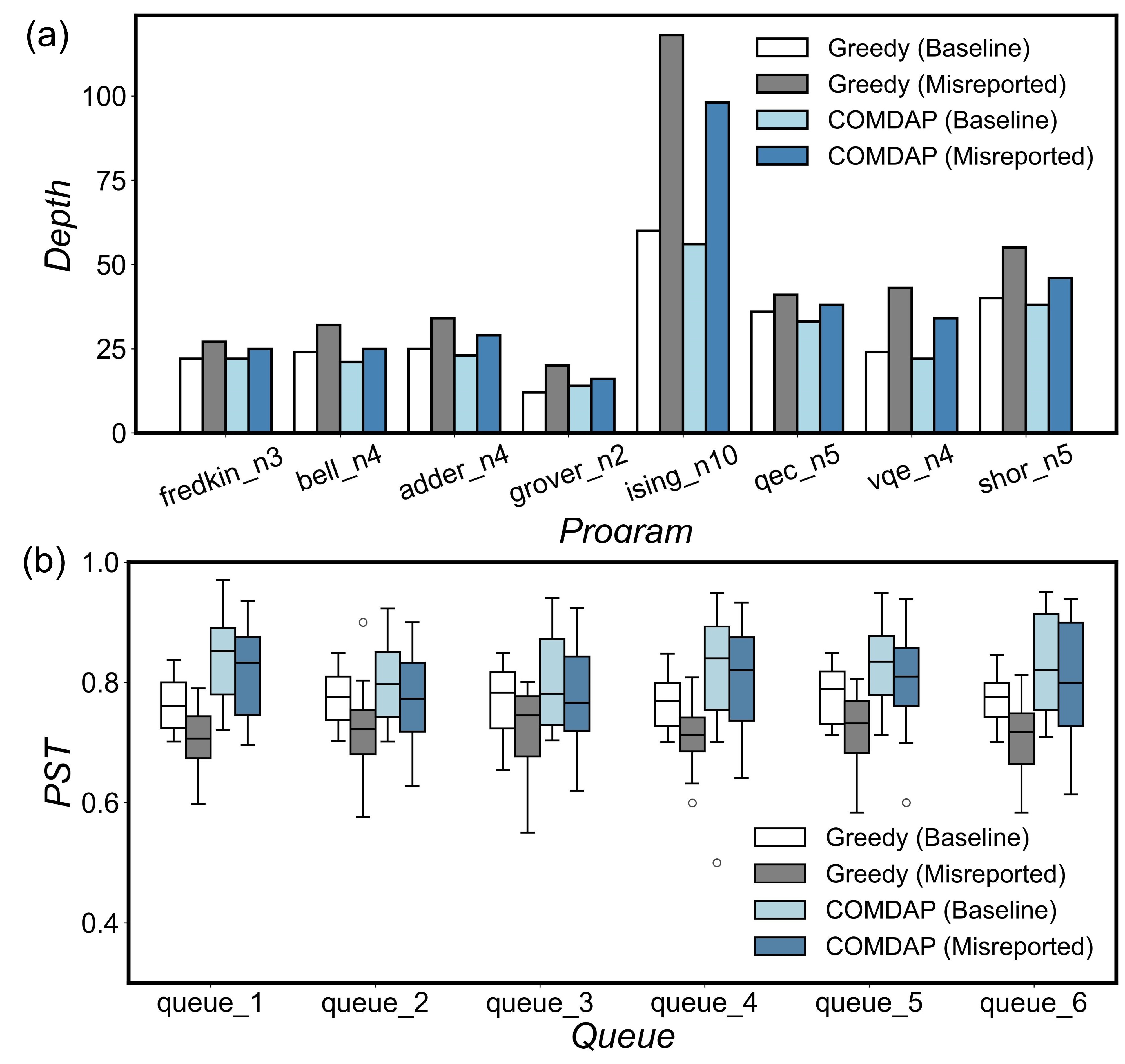}
  \caption{Effect of attack using Heuristic II on circuit depth and program fidelity. (a) Circuit depth of 8 different programs and (b) Probability of Successful Trials (PST) across different execution queues containing 40 programs each under baseline and misreported conditions. }
  \label{fig:fig4}
\end{figure}

We evaluated the impact of attack using Heuristic II by under-reporting the error rates of specific qubits on the Fake27QPulseV1 backend. Using Eq. (3), we identified $Q1$, $Q25$ and $Q14$ as high-connectivity qubits with the greatest minimum-path distance between them. Their error rates were under-reported by 15\%, 12\%, and 10\%, respectively. 
As shown in Fig. \ref{fig:fig4}(a), under adversarial conditions, the Greedy allocator exhibited an average 45\% increase in circuit depth across all programs. The COMDAP allocator showed more resilience, with an average 20\% increase in depth. 

We further extended our evaluation by running experiments on six different queues, each containing 40 programs. The values of \( n \), \( k_1 \), \( k_2 \), and \( k_3 \) were kept the same as in previous test. For each queue, we computed PST, which represents the fraction of trials yielding correct results—a higher PST indicates better execution fidelity. As shown in Fig. \ref{fig:fig4}(b), adversarial misreporting caused a drop in PST across all queues. The Greedy allocator experienced an average PST reduction of 7.8\%, while COMDAP showed a smaller drop of 3.5\%. The fidelity loss is directly linked to the increased circuit depth and SWAP operations from misreporting, which introduce additional noise.

The difference in circuit depth increase between Greedy and COMDAP stems from their allocation strategies. Greedy relies solely on individual qubit error rates ($E, R$) and connectivity ($d$), using CFM (as mentioned in section 2.3). Since CFM does not consider overall hardware structure, the misreported qubits appear more favorable, leading to inefficient mappings and excessive SWAP operations. In contrast, COMDAP employs a more holistic allocation strategy, using CRI. Along with error rates and connectivity, CRI also accounts for partition density ($D$) and compactness ($C$), making COMDAP less sensitive to local misreporting \cite{upadhyay2024share}. While it still experiences some degradation, its global optimization approach reduces the impact of manipulated error rates.


\section{Defense Strategy}
\label{sec:defense}
\begin{figure}[h]
  \centering
  \includegraphics[width=\linewidth]{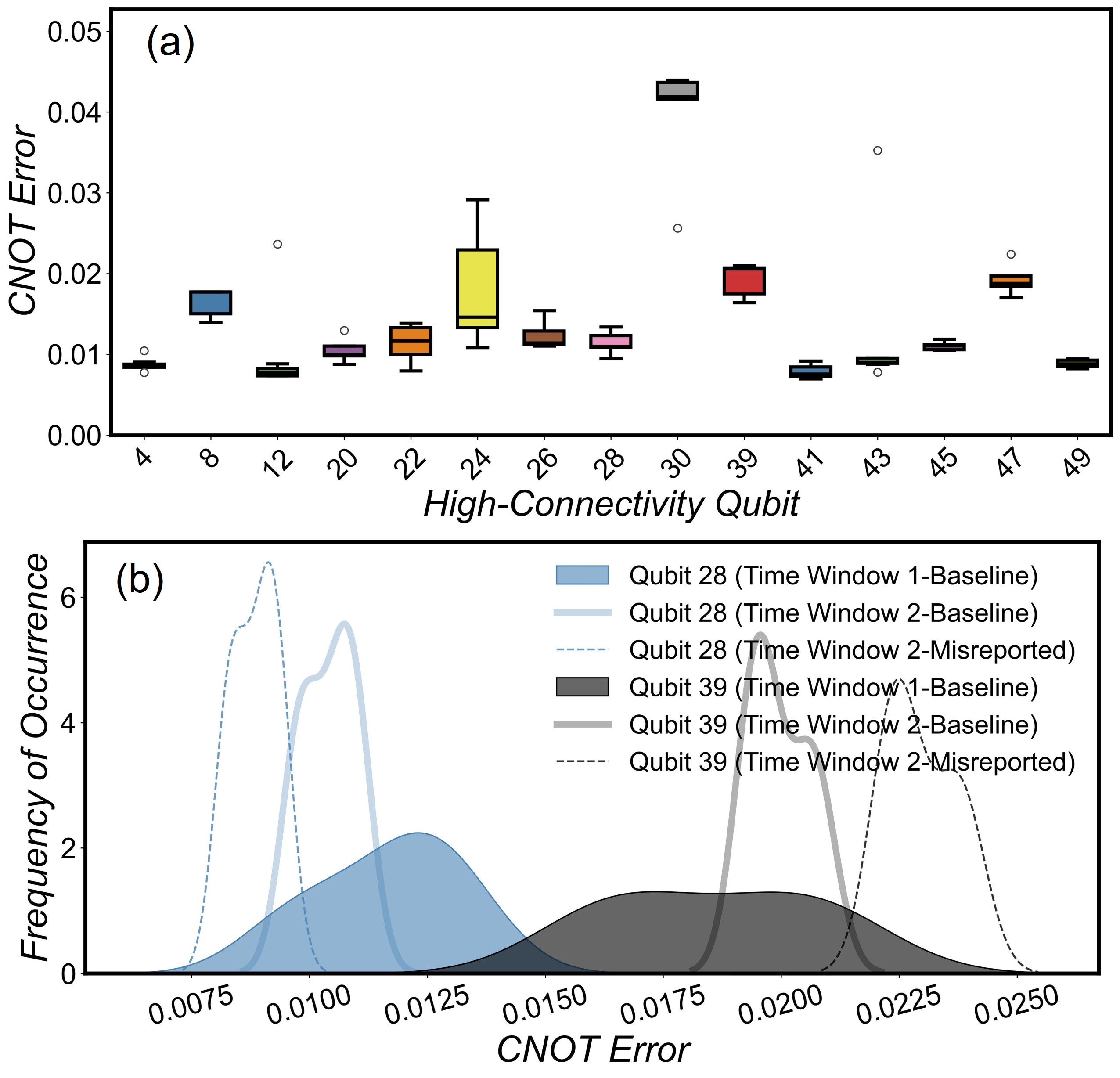}
  \caption{Temporal variation and effect of misreporting on CNOT error. (a) Natural CNOT error fluctuation for 15 high-connectivity qubits of $ibm\_kyiv$ over two weeks. (b) Comparison of baseline and misreported CNOT error distributions across two time windows.}
  \label{fig:fig5}
\end{figure}

Detecting adversarial misreporting is challenging due to temporal variation in error rates. If the inherent fluctuation in CNOT error is significant, small manipulations in reported values can remain undetected. To assess this challenge, we analyze the calibration data of the $ibm\_kyiv$ quantum processor over a two-week period. 
Our proposed attack heuristics target high-connectivity qubits, and in $ibm\_kyiv$, the maximum qubit degree is 3, so we focus on qubits with this connectivity. Since CNOT error is defined for links rather than individual qubits, we compute the average CNOT error for each qubit by taking the mean error of its three connected links. Fig. \ref{fig:fig5}(a) presents the temporal variation for 15 such qubits. The results show that CNOT error exhibits substantial natural variation over time, with an average fluctuation of 30.25\% across qubits. Both of our proposed heuristics misreport error rates by only $\pm$15\%, which is significantly lower than the observed natural fluctuation. As a result, threshold-based anomaly detection fails to distinguish misreporting from natural error drift. 

To overcome this, we propose a statistical divergence-based detection mechanism. Fig. \ref{fig:fig5}(b) illustrates the baseline and misreported CNOT error distributions for two high-connectivity qubits, 28 and 39, as an example. These qubits exhibited significant temporal variations of 40.67\% and 28.02\%, respectively, which prevents simple threshold-based anomaly detection. To generate Fig. \ref{fig:fig5}(b), we partition the calibration data into two one-week time windows and plot the Kernel Density Estimation (KDE) curves for the CNOT error distributions. The first time window serves as the initial baseline distribution. In the second time window, we compare two distributions: one derived from the actual calibration data (serving as an updated baseline) and another reflecting adversarial misreporting. The misreported error values are generated by randomly over-reporting or under-reporting CNOT errors by 15\%, following our proposed heuristics. We observe that the baseline distributions from the two time windows remain statistically similar, exhibiting only minor shifts in mean. However, the misreported distributions introduce a systematic bias, shifting the central tendency beyond the range of expected variation. This discrepancy, while imperceptible through threshold-based detection, is distinguishable using statistical divergence.

We leverage this insight to detect misreporting using statistical divergence metrics. By maintaining a historical record of error distributions for each qubit, we assess deviations using Kullback-Leibler (KL) divergence. KL divergence measures the difference between two probability distributions $P(x)$ and $Q(x)$ as:
\vspace{-3pt}
\[
D_{\text{KL}}(P || Q) = \sum_{x} P(x) \log \frac{P(x)}{Q(x)}
\]

where $P(x)$ represents the empirical distribution of historical error rates, and $Q(x)$ corresponds to the newly reported error distribution. A low divergence indicates natural temporal variation, whereas a high divergence signals potential misreporting. Although high temporal variation prevents immediate detection, misreporting alters the long-term statistical profile of error rates. By continuously evaluating KL divergence across calibration cycles, we can identify deviations unlikely to occur under natural fluctuations.

\section{Conclusions}
\label{sec:conclusions}
Accurate error calibration is essential for efficient resource allocation in multi-tenant quantum computing. However, reliance on third-party calibration services creates a vulnerability where adversaries can misreport error values without altering the physical hardware. We proposed an attack model based on strategic error misreporting and examined its impact on Greedy and COMDAP allocation frameworks. We further develop two heuristics to reduce hardware throughput and fidelity, respectively. We evaluate their impact using benchmark circuits on the Fake27QPulseV1 backend. Results show that adversarial misreporting misguides resource allocation frameworks, leading to inefficient qubit distribution, increased execution latency, and reduced computational fidelity. COMDAP is slightly more resilient than Greedy but still experiences significant degradation. Finally, high temporal variation in error rates makes threshold-based anomaly detection ineffective. To address this, we propose a statistical detection framework that analyzes long-term deviations in reported error distributions across calibration cycles. 
\begin{acks}
The work is supported in parts by the National Science Foundation (NSF) (CNS-1722557, CCF-1718474, OIA-2040667, DGE-1723687 and DGE-1821766).
\end{acks}

\bibliographystyle{ACM-Reference-Format}
\bibliography{sample-base}

\appendix









\end{document}